\pdfoutput=1
\documentclass[superscriptaddress,aps,preprintnumbers,amsmath,amssymb,prl,nofootinbib, reprint]{revtex4-2}
\usepackage{graphicx}
\usepackage{bm}
\usepackage{color}
\usepackage{amsmath,amssymb}	
\usepackage{setspace}
\usepackage{longtable}
\usepackage{booktabs}
\usepackage{comment}
\usepackage{array}
\usepackage{cancel}
\usepackage{enumitem}
\usepackage{todonotes}
\usepackage{physics}
\usepackage{siunitx}

\usepackage[subpreambles, sort]{standalone}
\usepackage{hyperref}

\def\slashchar#1{\setbox0=\hbox{$#1$} % set a box for #1
\dimen0=\wd0 % and get its size
\setbox1=\hbox{/} \dimen1=\wd1 % get size of /
\ifdim\dimen0>\dimen1 % #1 is bigger
\rlap{\hbox to \dimen0{\hfil/\hfil}} % so center / in box
#1 % and print #1
\else % / is bigger
\rlap{\hbox to \dimen1{\hfil$#1$\hfil}} % so center #1
/ % and print /
\fi}

\setuptodonotes{inline}
\newcommand{\newton}{G_{\mathrm{N}}}

\newcommand{\heading}[1]{\textit{#1}---}
\begin{document}
\newcolumntype{Y}{>{\centering\arraybackslash}p{23pt}} 

%%%%%%%%%%%%%%%%%%%%%%%%%%%%%%%%%%
%%%%%%%%%%% Title page %%%%%%%%%%%
%%%%%%%%%%%%%%%%%%%%%%%%%%%%%%%%%%

\preprint{IPMU25-0039}

\title{Do Cosmic String Segments Emit Gravitational Waves?}

\author{Akifumi Chitose}
%\email[e-mail: ]{achitose@icrr.u-tokyo.ac.jp}
\affiliation{ICRR, The University of Tokyo, Kashiwa, Chiba 277-8582, Japan}

\author{Masahiro Ibe}
%\email[e-mail: ]{ibe@icrr.u-tokyo.ac.jp}
\affiliation{ICRR, The University of Tokyo, Kashiwa, Chiba 277-8582, Japan}
\affiliation{Kavli Institute for the Physics and Mathematics of the Universe (WPI), \\The University of Tokyo Institutes for Advanced Study, \\ The University of Tokyo, Kashiwa 277-8583, Japan}

\author{Shunsuke Neda}
%\email[e-mail: ]{neda@icrr.u-tokyo.ac.jp}
\affiliation{ICRR, The University of Tokyo, Kashiwa, Chiba 277-8582, Japan}

\author{Satoshi Shirai}
%\email[e-mail: ]{satoshi.shirai@ipmu.jp}
\affiliation{Kavli Institute for the Physics and Mathematics of the Universe 
(WPI), \\The University of Tokyo Institutes for Advanced Study, \\ The University of Tokyo, Kashiwa 277-8583, Japan}

\date{\today}
\begin{abstract}
Cosmic strings are predicted in various extensions of the Standard Model, including grand unified theories.
Depending on the symmetry-breaking pattern, they can be either topologically stable or metastable.
Intriguingly, metastable strings have been proposed as a possible origin of the gravitational wave (GW) background observed by recent pulsar timing array  experiments.
When metastable strings decay, they fragment into segments with monopoles and antimonopoles attached at their endpoints.
The monopole and antimonopole are strongly pulled by the string tension.
Violent oscillations of these segments have been considered as a potential GW source, in addition to contributions from string loops.
We show that, in realistic situations, the monopoles frequently collide with thermal fluctuations on the string segments, which act as a resistance and prevent the oscillation.
As a result, we find that the contribution from string segments to the GW background is negligible.
\end{abstract}

\maketitle
%%%%%%%%%%%%%%%%%% Section %%%%%%%%%%%%%
\heading{Background}%
\label{sec:intro}%
%%%%%%%%%%%%%%%%%%%%%%%%%%%%%%%%%%%%%%%%
Recently, multiple pulsar timing array (PTA) collaborations have reported evidence for a stochastic gravitational wave (GW) background in the nanohertz band~\cite{NANOGrav:2023gor,EPTA:2023fyk,Reardon:2023gzh,Xu:2023wog}, drawing renewed attention to cosmic strings as a compelling source of such signals.
Cosmic strings are topological defects formed during symmetry-breaking phase transitions in the early Universe, specifically when a symmetry group $G$ is spontaneously broken to a subgroup $H$ such that $\pi_1(G/H) \neq 0$.
Once formed, a cosmic string network evolves and radiates energy predominantly via GWs (see, e.g., Ref.\,\cite{Vilenkin:2000jqa}).

While ordinary cosmic strings are stable topological defects associated with nontrivial $\pi_1(G/H)$, current PTA results intriguingly favor \emph{metastable} strings. These typically arise in multi-step symmetry breaking $G \to G' \to H$ where $\pi_1(G/H) = 0$ but $\pi_1(G'/H) \neq 0$\,\cite{Vilenkin:1982hm} (see Ref\,\cite{Chitose:2025cmt} for a more extended symmetry breaking pattern). In such cases, strings formed at the $G' \to H$ transition are unstable and decay via pair production of monopoles associated with nontrivial $\pi_2(G/G') \simeq \pi_1(G')/\pi_1(G)$~\cite{Preskill:1992ck}. Their late-time decay reproduces the observed blue-tilted GW spectrum~\cite{Leblond:2009fq,Buchmuller:2019gfy,Buchmuller:2020lbh,Buchmuller:2021mbb,Buchmuller:2023aus}.

The string decay rate per unit length is estimated as~\cite{Preskill:1992ck}
\begin{align}
\Gamma_d = \frac{\mu_\mathrm{s}}{2\pi} e^{-\pi \kappa} \ , \quad
\sqrt{\kappa} \simeq \frac{M_m}{\sqrt{\mu_\mathrm{s}}} \ ,
\label{eq:rootkappaPreskill}
\end{align}
within the so-called thin-wall approximation (see also Ref.\,\cite{Chitose:2023dam}).
Here, $M_m$ and $\mu_\mathrm{s}$ denote the monopole mass and the string tension, respectively.
The PTA data favor $G_\mathrm{N} \mu_\mathrm{s} \sim 10^{-5}$ and $\sqrt{\kappa} \sim 8$~\cite{NANOGrav:2023hvm}, where $G_\mathrm{N}$ is Newton constant.
When the Hubble expansion rate drops to $H \simeq \sqrt{\Gamma_d}$, long strings decay into segments with monopoles and antimonopoles attached to the endpoints.
For the metastable strings favored by the PTA signal, this decay occurs at a cosmic temperature of $T = \order{0.1}$\,MeV.

String segments potentially contribute significantly to the GW spectrum. 
Indeed, the analysis by NANOGrav\,\cite{NANOGrav:2023hvm} distinguishes two scenarios depending on whether the segment contributions are included (META-LS) or excluded (META-L). 
These correspond to segments without and with unconfined magnetic flux from the monopoles at the endpoints, respectively.
The latter dissipate energy primarily through particle emission rather than GWs \cite{Berezinsky:1997kd}.

In Ref.\,\cite{NANOGrav:2023hvm}, the GW emission from string segments is estimated based on monopole oscillations driven by the string tension. 
The power spectrum of the emitted GW scales as $k^{-1}$, where $k$ denotes the mode number 
of the GW with frequency $2\pi k/L_\mathrm{s}$, for a segment of length $L_\mathrm{s}$~\cite{Martin:1996cp}. 
For straight segments, the upper cutoff of the $k^{-1}$ spectrum is set by $k_\mathrm{max} \sim \gamma_m^2$ where
$\gamma_m$ is
the maximum boost factor of the endopoint monopole.
However, the GW contribution from the segment
remains uncertain. 
For example, a conventional choice of $k_\mathrm{max}$ is $10^5$~\cite{Buchmuller:2021mbb}, which yields a GW power comparable to that from loops, 
although the precise value of $k_\mathrm{max}$ is still under debate~\cite{Servant:2023tua}.
Given the upcoming PTA~\cite{Carilli:2004nx, Janssen:2014dka, Weltman:2018zrl}  and interferometer experiments~\cite{LISA:2017pwj, Baker:2019nia, Blanco-Pillado:2024aca, Abac:2025saz}, which will enable unprecedented tests of the stochastic GW background from cosmic strings, precise predictions of the GW signals are crucial.

In this \textit{Letter}, we argue that such oscillations do not actually occur. 
In particular, we show that collisions of the monopole with thermal fluctuations on the string counteract the monopole acceleration. 
We demonstrate that, especially in the parameter region favored by the PTA GW signal, the monopole ceases to accelerate well before the segment shrinks by an amount comparable to the Hubble length. 
As a result, we find that segments do not contribute to the GW spectrum in comparison to long string loops.

%%%%%%%%%%%%%%%%%% Section %%%%%%%%%%%%%
\heading{Segment dynamics}%
\label{sec:EOM}%
%%%%%%%%%%%%%%%%%%%%%%%%%%%%%%%%%%%%%%%
Here we review the dynamics of segments following Ref.~\cite{Martin:1996cp}.
The segment forms a two-dimensional worldsheet parameterized by $(\zeta^0,\zeta^1)$ in the four-dimensional spacetime,
\begin{align}
    x^\mu(\zeta^a) \ , \quad \mu = 0, 1, 2, 3\ , \quad a = 0,1\ .
\end{align}
The endpoint monopole ($m$) and the antimonopole ($\overline{m}$)
follow worldlines on the string worldsheet
\begin{align}
    x_i^\mu = x^\mu(\zeta^0,\sigma_i(\zeta^0)) \quad i = m, \overline{m}\ ,  
\end{align}
The action is given by
\begin{align}
    S=&-\sum_{i=m,\overline{m}}M_m\!\int\! d\zeta^0\sqrt{-g_{\mu\nu}\dot{x}_i^\mu \dot{x}_{i}^\nu}
    \cr
   & -\mu_\mathrm{s} 
    \int\! d\zeta^0\int_{\sigma_m(\zeta^0)}^{\sigma_{\overline{m}}(\zeta^0)}d\zeta^1 \sqrt{\det(-g_{\mu\nu}x^{\mu}{}_{,a} x^{\nu}{}_{,b})} \ , 
\end{align}
where $x^\mu{}_{,a} := \partial x^\mu/\partial \zeta^a$, $\dot{x}^\mu_i:=d x^\mu/d\zeta^0$ and $g_{\mu\nu}$ is the spacetime metric.

In the flat background $x^\mu=(t,x,y,z)$,
we fix the transverse traceless gauge and choose $\zeta^0=t$.
The temporal component of the equation of motion for the monopole is
\begin{align}
    \dot{\gamma}_{m} M_m
    =     \mu_\mathrm{s} \dot{\sigma}_{m}\ .
\end{align}
When a string segment shortens by an invariant length $\ell$, the attached magnetic monopole is accelerated to
\begin{align}
\label{eq:gamma length}
\gamma_m(\ell) = \frac{E_m}{M_m} = \frac{\mu_\mathrm{s}}{M_m} \times
\ell\ 
\end{align}
where $E_m$ is the monopole’s energy obtained from the string shrinking.

If the string segment is perfectly straight, the monopole at the string endpoint experiences no resistance and accelerates until it meets the other end, leading to oscillation.
In this case, the boost factor of the monopole gained over a Hubble time is
$\gamma_m(H^{-1})$.
For monopole mass $M_m \sim \sqrt{\mu_\mathrm{s}} \sim 10^{16}$\,GeV which decays at the cosmic temperature of $T = \order{1}$\,MeV, for instance, the boost factor reaches
\begin{align}
\label{eq:gamma_M40}
\gamma_m \simeq 10^{40} \ .
\end{align}
Such an enormous boost factor implies that the monopole carries an energy comparable to the solar mass.
This raises several concerns, including the validity of the effective theory describing the monopole.

%%%%%%%%%%%%%%%%%% Section %%%%%%%%%%%%%
\heading{Fluctuations on cosmic strings}%
\label{sec:thermal}%
%%%%%%%%%%%%%%%%%%%%%%%%%%%%%%%%%%%%%%%%
%%%%%%%%%%%%%%%%%%%%%%%%%%%%%%%%%%%%%%%%
\begin{figure}[t]
    \centering
    \includegraphics[ width=.4\textwidth ]{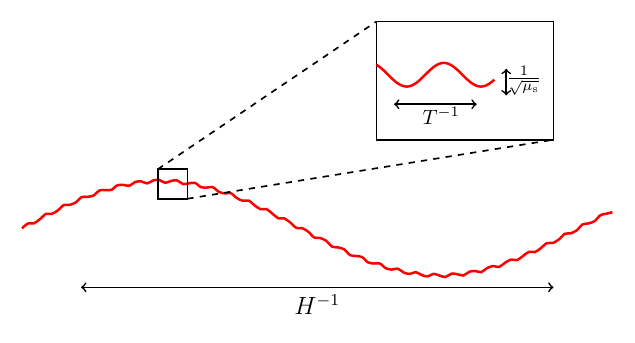}
    \caption{Schematic illustration of a cosmic string, showing curvature on the Hubble scale and thermal-scale fluctuations with a temperature $T$.
    The typical wave length is $T^{-1}$ 
    with a typical amplitude
    of $\mathcal{O}(\mu_\mathrm{s}^{-1/2})$
    in the zoomed-in region.}
    \label{fig:fluctuation}
\end{figure}
The curvature radius of a string or string segment on cosmological scales is typically of order $H^{-1}$.
Besides, transverse thermal fluctuations with much shorter length scale also exist (see Fig.\,\ref{fig:fluctuation}). 
To discuss their effects, we focus on a string section of an intermediate length scale.
We approximate it
as a straight string along the $z$-axis. 
The worldsheet coordinates are taken as $(\zeta^0,\zeta^1)=(\eta,z)$,
where $\eta$ is the conformal time.
When the amplitude of the transverse fluctuations is sufficiently small compared to the curvature radius, the Nambu-Goto action leads to a two-dimensional massless field,
\begin{align}
    S\supset  
    &- \frac{\mu_\mathrm{s}}{2}\!
     \int \! d\eta\, dz\, a(\eta)^2 \qty[\qty(\pdv{\vec{x}_\perp}{\eta})^2\! - \qty(\pdv{\vec{x}_\perp}{z})^2 ]\ ,
\end{align}
where $a(\eta)$ is the scale factor of the Universe, and 
$\vec{x}_\perp(\eta,z)$ represents the transverse fluctuations.
Hereafter we assume the radiation-dominated Universe where $a(\eta)\propto \eta$.

Now, let us discuss the number density of zero modes present on a long string. 
After the string formation in the very early Universe, 
the system settles into a thermal plasma containing a few long cosmic strings per Hubble volume. 
In this setting, it is reasonable to assume that the zero modes follow a thermal distribution governed by the Bose-Einstein statistics,
\begin{align}
    f(p_\mathrm{fr}) \simeq \frac{1}{e^{p_\mathrm{fr}/T_\mathrm{fr}} - 1} \ ,
\end{align}
where $T_\mathrm{fr} \lesssim \sqrt{\mu_\mathrm{s}}$ 
is the temperature below which thermal friction on the strings becomes irrelevant compared to the expansion rate. 
If the interaction between the string and the thermal bath is weak, finer fluctuations may remain on the long string. 
Such situations are expected in models where cosmic strings are formed at the end of inflation, as in the case of hybrid inflation~\cite{Linde:1991km,Linde:1993cn} (see, e.g., Ref.\,\cite{Chitose:2024pmz}).
In such cases, more transverse modes reside on the string, and  the deceleration of the monopole can be even more significant.
Therefore, the analysis that follows is conservative.

At later times, the fluctuation modes are red-shifted and the distribution evolves into
\begin{align}
    f(p) \simeq \frac{1}{e^{p/T_\mathrm{s}} - 1} \ ,\quad T_\mathrm{s} := \frac{a_\mathrm{fr}}{a} T_\mathrm{fr} \ ,
\end{align}
where $a_\mathrm{fr}$ is the scale factor 
at $T_\mathrm{s} \simeq T_\mathrm{fr}$.
In the present analysis, we use the Standard Model (SM) equation of state to track the evolution of the scale factor \cite{Saikawa:2018rcs,*Saikawa:2020swg}.
Accordingly, the number density of zero modes per unit string length is given by
\begin{align}
    n = 2\int \frac{dp}{2\pi} f(p)
      = 2T_\mathrm{s} \int \frac{dx}{2\pi} \frac{1}{e^x - 1}
      \simeq \frac{T_\mathrm{s}}{\pi} \ln\frac{M_\mathrm{Pl}}{T_\mathrm{fr}} \ , \label{eq:zeromodedensity}
\end{align}
where the factor of two counts the transverse dimensions and $M_\mathrm{Pl}=(8\pi \newton)^{-\frac{1}{2}}$.
We have also assumed that the infrared (IR) cutoff of the integral is determined by
\begin{align}
    \frac{p_\mathrm{fr}}{T_\mathrm{fr}} \simeq \frac{T_\mathrm{fr}}{M_\mathrm{Pl}} \ ,
\end{align}
although the following discussion is not sensitive to the precise value of the IR cutoff.

Incidentally, the power of GW emission per unit length from the transverse fluctuations is of order $T_\mathrm{s}^5/M_\mathrm{Pl}^2$. Therefore, the energy loss of the fluctuations
by the GW emission is negligible compared with the red-shift due to the cosmic expansion.

%%%%%%%%%%%%%%%%%% Section %%%%%%%%%%%%%
\heading{Limit on monopole acceleration}%
\label{sec:upper_lim}%
%%%%%%%%%%%%%%%%%%%%%%%%%%%%%%%%%%%%%%%%
Now, let us discuss the
upper limit on the energy of
the monopole.
Consider a long straight cosmic string with the Hubble length scale, $H^{-1}$. 
We assume that this string decays into segments when $H \sim \Gamma_d^{1/2}$.
Once the segment shortens by cosmological length $\ell$, 
the monopole becomes relativistic:
\begin{align}
    \gamma_m = \frac{\mu_\mathrm{s}}{M_m} \times \ell \gg 1 \ .
\end{align}
However, it collides with the zero modes,
transferring its momentum.
The center-of-mass energy is given by
\begin{align}
    \sqrt{s} \simeq 2\sqrt{E_m T_\mathrm{s}} = 2\sqrt{\gamma_m M_m T_\mathrm{s}} \ .
\end{align}
Accordingly, after one collision, the monopole loses its energy by
\begin{align}
   \mathit{\Delta} E &\simeq    \frac{\left(\gamma_m +\sqrt{\gamma_m^2-1}\right) M_m T_\mathrm{s}}{\left(\gamma_m-\sqrt{\gamma_m^2-1}\right)
   M_m+2 T_\mathrm{s}}\ , 
\end{align}
measured in the rest frame of the thermal bath.

\begin{figure}[t]
    \centering
    \includegraphics[width=.4\textwidth ]{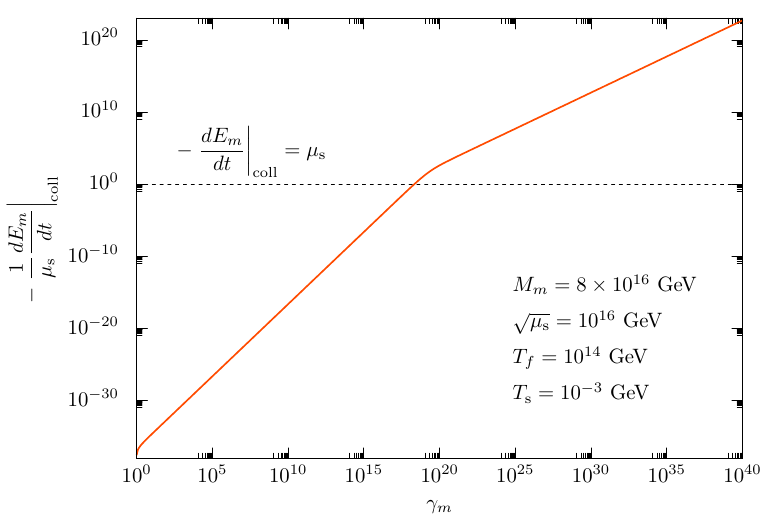}
    \caption{
        The monopole energy loss rate divided by the drag force. In this figure, we fix the monopole mass $M_m$, the string tension $\mu_\mathrm{s}$, the temperature of the transition to friction-free strings $T_f$ and the string temperature at the decay $T_\mathrm{s}$.
    }
    \label{fig:gammaCR}
\end{figure}

Such collisions occur at a rate of $n\times(1+\beta_m)$ per unit time for the monopole moving in the speed of $\beta_m$. Therefore, the energy loss rate of the monopole is
\begin{align}
   \frac{dE_m}{dt}\bigg|_{\mathrm{coll}} &\simeq  -\frac{(1+\beta_m)^2\gamma_m M_m T_s^2 \ln \frac{M_\mathrm{Pl}}{T_{\mathrm{fr}}}}{\pi 
   \left(2 T_s+\left(1-\beta_m \right)\gamma_m M_m \right)} \ .
   \label{eq:dEdtcollison}
\end{align}
Figure~\ref{fig:gammaCR} shows
$dE_m/dt|_\mathrm{coll}$ as a function
of the boost factor $\gamma_m$.
Here, we take the monopole mass and string tension to be around $10^{16}\,\mathrm{GeV}$, as favored by the PTA results.
It demonstrates that the $dE_m/dt|_\mathrm{coll}$ exceeds string tension at $\gamma_m\sim 10^{20}$.
%%%%%%%%%%%%%%%%%% Figure %%%%%%%%%%%%%
%%%%%%%%%%%%%%%%%%%%%%%%%%%%%%%%%%%%%%%
When the energy loss rate \eqref{eq:dEdtcollison} balances the string tension $\mu_\mathrm{s}$, the monopole stops accelerating. 
This occurs when
\begin{align}
\label{eq:critical}
\gamma_m^{\mathrm{(cr)}}
\simeq
\max{\left[\frac{\mu_\mathrm{s}}{M_m T_\mathrm{s}},~\frac{\sqrt{\mu_\mathrm{s}}}{T_\mathrm{s}}\,\right]}\ .
\end{align}
Notice that this also gives the upper limit on the center-of-mass energy of the collision,
\begin{align}
    \sqrt{s^{(\mathrm{cr})}} \simeq \max[\,\mu_\mathrm{s}^{1/2},(\mu_{\mathrm{s}}M_m^2)^{1/4}\,]\ .
\end{align}
Thus, for metastable strings with $\sqrt{\kappa} > \order{1}$, it does not exceed the monopole mass scale, indicating that the effective theory of the monopole remains valid.

\begin{figure}[t]
    \centering
\includegraphics[width=.4\textwidth]{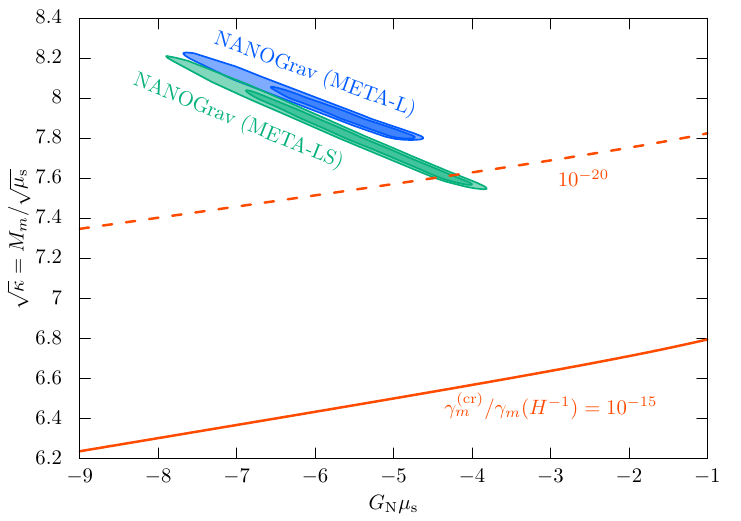}
    \caption{Contour plot of $\gamma_M^{(\mathrm{cr})}/\gamma_M(H^{-1})$.
     In the whole parameter region in this figure, the string segments do not oscillate.}
    \label{fig:result}
\end{figure}

Figure~\ref{fig:result} shows the ratio between the critical boost factor \eqref{eq:critical}
with the na\"ive one \eqref{eq:gamma length} with $\ell=H^{-1}$
in the parameter space of $(\newton\mu_\mathrm{s}, \sqrt{\kappa})$.
In the figure, we assume the
sudden 
decay at $H = \sqrt{\Gamma_d}$.
Figure shows that 
$\gamma_m^\mathrm{(cr)}/\gamma_m(H^{-1})\ll 1$ in the entire parameter space.
In the figure, we also show the posterior distributions of
($\newton\mu_\mathrm{s}$, $\sqrt{\kappa}$)
favored by the PTA signal reported in Ref.\,\cite{NANOGrav:2023hvm}
for the META-L and the META-LS scenarios.

In the string network, string segments typically have curvature radii of the Hubble length. 
Nevertheless, our argument remains valid even for such curved strings, 
since monopole deceleration only requires straightness over a length scale 
$\mathit{\Delta} \ell \sim M_m/\mu_\mathrm{s} \times \gamma_{m}^{(\mathrm{cr})}$, 
which is much shorter than the Hubble length, i.e., 
$\mathit{\Delta} \ell/H^{-1} \simeq \gamma_{m}^{(\mathrm{cr})}/\gamma_m(H^{-1}) \ll 1$ (see Fig.\,\ref{fig:result}). 
As a result, we find that the string segments formed from the string network do not enter the oscillation regime, 
and instead shrink at a constant velocity down to a length of order $\mathit{\Delta} \ell$. 
Since the monopole also loses angular momentum, 
configurations such as rotating segments with cosmological length scales, 
as discussed in Refs.\,\cite{Martin:1996cp,Babichev:2004gy}, are not expected.

Note that the deceleration due to collisions with zero modes remains relevant even in the presence of monopole energy loss via massless particle radiation associated with unconfined flux~\cite{Berezinsky:1997kd}. 
Since the radiation power is given by the Larmor formula~\cite{Berezinsky:1997kd} and is approximately proportional to $\dot{\gamma}_m^2$, the monopole would undergo a uniform acceleration of $\order{\sqrt{\mu_\mathrm{s}}}$ if such collisions were absent. 
Therefore, the monopoles are decelerated primarily by the collisions even in this case.

\heading{Fate of string segments}%
Now, let us discuss how the initial energy
of a segment, $\mu_\mathrm{s} H^{-1}$,
is released. 
Given that the center-of-mass energy for a monopole–zero mode collision reaches $\sqrt{s^{(\mathrm{cr})}} \sim 
\sqrt{\mu_\mathrm{s}}$, inelastic processes like 
\begin{align}
&\text{monopole} + \text{zero mode} \cr 
&\phantom{XXX}
\to 
\text{monopole} + \text{massive modes}
\end{align}
are triggered, where
the massive modes include some 
massive gauge boson/Higgs boson excitations. 
Such massive modes eventually decay into 
light particles which may or may not include SM particles. 
Since these decays occur on microscopic timescales, 
the monopole energy 
is eventually radiated into the environment.

Let us comment on the effects of segment energy dissipating into the thermal bath.
Metastable strings favored by the PTA signal are expected to decay at a cosmic temperature of $T = \order{0.1}$\,MeV.
If the segment energy is transferred to SM particles, the resulting late-time energy injection is tightly constrained, as it can modify the primordial light element abundances from Big Bang Nucleosynthesis.
The yield of injected SM particles with energy $E_{\mathrm{in}}$ is estimated as
\begin{align}
E_{\mathrm{in}} Y_\mathrm{in} \sim G_\mathrm{N}\mu_\mathrm{s} T
\sim 10^{-8}
\qty(\frac{ G_\mathrm{N}\mu_\mathrm{s}}{10^{-5}})
\qty(\frac{T}{\mathrm{MeV}}) \ .
\label{eq:EY}
\end{align}
Here, we assume that the string network energy density was in the scaling regime just before the decay, $\rho_\mathrm{s} \sim G_\mathrm{N}\mu_\mathrm{s} \times \rho_R$, with $\rho_R$ the radiation energy density.
Comparing with the constraints in Ref.\,\cite{Kawasaki:2017bqm}, the estimated energy injection \eqref{eq:EY} is below the existing bounds.
However, effects such as the injection of ultrahigh-energy particles into the plasma and spatial anisotropy of the injection, which are not included in current analyses, remain to be clarified.

%%%%%%%%%%%%%%%%%% Section %%%%%%%%%%%%%
\heading{Discussions}%
\label{sec:disc}%
In this work, we have studied the dynamics of monopoles at the endpoints of cosmic string segments in the context of metastable cosmic strings relevant to the PTA GW signal. 
We have shown that transverse thermal fluctuations on strings counteract the monopole acceleration through frequent scatterings. 
As a result, the monopole energy gain saturates at a critical boost factor $\gamma_m^{\mathrm{(cr)}}$, which is significantly smaller than the na\"ive estimate assuming uninterrupted acceleration over a Hubble time. 
In particular, for parameter values favored by current PTA data, we find that monopoles do not enter the oscillatory regime, and most of the segment energy is dissipated non-gravitationally, rendering the resulting GW emission negligible. 
This conclusion holds whether or not the endpoint monopoles carry unconfined flux.

We have focused on the impact of the endpoint monopole dynamics on the GW signal. However, our results may also have implications for the spectrum of ultra-high-energy cosmic rays originating from the endpoint monopoles~\cite{Berezinsky:1997kd,Berezinsky:1998ft}.

To fully understand the ultimate fate of the string segments,
however, our perturbative analysis is not sufficient.
For instance, within our treatment, segments potentially become black holes, as the center-of-mass energy of the monopole-antimonopole pair is $2\gamma_m^{\mathrm{(cr)}}M_m$, whose Schwarzshild radius greatly exceeds the typical length scale of annihilation $M_m^{-1}$.
However, once the monopole reaches such a high energy, our current perturbative approach to evaluating its energy loss becomes unreliable.
In reality, the fate of the segments may be determined by non-perturbative dynamics, such as microscopic string loop formations.
These effects are expected to substantially enhance the energy dissipation, resulting in a final monopole energy far smaller than the present estimate.

%Eventual violent motion of the monopoles and the string is expected to prevent $\gamma_m$ from growing and/or lead to efficient energy dissipation by, e.g., loop formation.

%However, once the monopole reaches such a high energy, our current perturbative approach to evaluating its energy loss becomes unreliable.
%Such effects are expected to substantially enhance the dissipation, resulting in a final monopole energy far smaller than the present estimate.

\begin{acknowledgments}
\heading{Acknowledgments}
This work is supported by Grant-in-Aid for Scientific Research from the Ministry of Education, Culture, Sports, Science, and Technology (MEXT), Japan,  22K03615, 24K23938 (M.I.), and by World Premier International Research Center Initiative (WPI), MEXT, Japan. 
The work of S.S. is supported by DAIKO FOUNDATION. 
This work is also supported by Grant-in-Aid for JSPS Research Fellow 
JP24KJ0832 (A.C.) and JP25KJ1030 (S.N.)
This work is supported by FoPM, WINGS Program, the University of Tokyo (A.C.).
This work is supported by JST SPRING, Grant Number JPMJSP2108 (S.N.).
M.I. would like to thank Dr.~Y.~Kanda for useful discussions.
\end{acknowledgments}

\bibliography{bibtex}

\end{document}